# Biological Crowding Annihilates Terahertz Transmission Nonlinearity in Aqueous Protein Solutions

Ellen M. Adams[1,2]*, Igor Ilyakov[3], Manthan Raj[1,2], Daniel Dornbusch[1,2], Thales V. A. G. de Oliveira[3], Atiqa Arshad[3], Gulloo Lal Prajapati[3], Alexey Ponomaryov[3], Jan-Christoph Deinert[3]

[1]*Institute of Resource Ecology, Helmholtz Zentrum Dresden Rossendorf, 01328 Dresden, Germany*
[2]*Cluster of Excellence Physics of Life, Technische Universität Dresden, 01307 Dresden Germany*
[3]*Institute of Radiation Physics, Helmholtz Zentrum Dresden Rossendorf, 01328 Dresden, Germany*

**\* Correspondence:**
e.adams@hzdr.de

**Abstract**

Hydration water is vital for the stabilization of protein structure and function. The strong interaction of hydration water with the protein surface brings into question how dynamics and asymmetry of hydrogen bonds are perturbed for hydration water compared to bulk water. Here, z-scan transmission measurements at 0.5 Terahertz (THz) were performed for dilute and concentrated lysozyme solutions. A giant nonlinear absorption coefficient was found for dilute lysozyme solutions that is ten times greater than previous studies. This giant nonlinear response stems from the high average THz power generated by the TELBE free electron laser source, which drives the formation of a persistent thermal lens. In contrast, concentrated lysozyme solutions did not demonstrate a nonlinear response, revealing that crowding annihilates the thermal lensing effect. These results indicates that the THz nonlinear transmission of aqueous proteins solutions depends on the amount of hydration water present, and opens to the door to understanding the nonlinear optical properties of biologically relevant systems.

**Introduction**

Water is essential for biological function of proteins, where protein conformation, stability and reactivity are impacted by water molecules in the hydration shell.[1] Hydration water has perturbed physicochemical properties relative to bulk water, including a preferred spatial arrangement, reduced hydrogen bonding strength and slowed water dynamics.[2,3] The exact thickness of the hydration shell remains debated in literature,[4] as a variety of techniques report different length scales.[5–8] Water interactions are cooperative rather than additive, and therefore predicting how a single point mutation in a protein sequence impacts the hydration water network, or conversely how solvent conditions influence protein conformation becomes challenging. Cooperative effects are well demonstrated by the Hofmeister series, where the ordering of chaotropic and kosmotropic solutes on protein stability is unique for specific proteins.[9,10] Such effects are thought to be driven by ion-protein and water-protein interactions that perturb the protein hydration shell,[11] and demonstrate that disentangling the relationship between protein driven and water driven dynamics is complex.

Extensive theoretical and experimental studies have been carried out over the years to gain a better understanding of water at the protein interface.[12–15] Labeling protein surfaces with solvent exposed probes has allowed techniques such as EPR and time dependent stokes shift to provide insight into the site specific local environment, where the microviscosity, dielectric constant, and number of hydrogen bonds can be obtained.[16,17] Ultrafast time resolved methods have been applied to investigate the structure and dynamics of both proteins and aqueous solutions,[18,19] and the targeted excitation of specific molecular groups has provided insight into the relationship of protein motions and solvent fluctuations.[20] For instance, investigation of globular protein surfaces labeled with metal carbonyls revealed that hydration water has a fast decay response after IR excitation, while fluctuations of the protein occurred at much slower time scales.[21] Dynamics of the molecular reorientation of hydration water determined from optical Kerr effect (OKE) studies found that water in the first hydration shell is slowed down to a greater extent for proteins with hydrophilic surfaces compared to those with hydrophobic surfaces.[22] Similarly, Terahertz (THz) spectroscopy has shown that steady state dynamics of the long range hydration shell is correlated to the local surface electrostatic potential of folded proteins.[23]

In the last twenty years THz radiation has emerged as a powerful tool to elucidate solvation properties.[24] THz radiation is resonant with low frequency motions of the water hydrogen-bonding network and is sensitive to perturbations in water solvation dynamics. THz absorption measurements have identified spectroscopic signatures of hydration water for both ions and proteins,[25,26] which have been similarly observed in THz Raman spectral densities obtained from OKE measurements.[22,27,28] Intense THz sources induce a strong anisotropy in water that is 3-6 times greater than in the optical or IR regimes,[29,30] and has opened the door for ultrafast and nonlinear methods to probe THz driven phenomena in aqueous solutions. For example, THz pump – optical probe experiments have been performed to study the Kerr effect in water, revealing that energy quickly dissipates from rotational modes to translational modes of the hydrogen bonding network.[31] This energy transfer is impacted by ions, where cations enhance the effect and anions suppress it.[32] Such behavior resembles the Hofmeister effect, and suggests that the ability of ions or osmolytes to stabilize protein structure is linked to their hydration shell. Investigation of protein hydration with these advanced techniques is lacking, despite a broad interest in developing knowledge of THz driven phenomena in biological systems. For instance, recent dielectric relaxation measurements found that THz irradiation perturbs hydration water dynamics of lysozyme and causes an increase in the number of hydrogen bonds in the hydration shell, resulting in more hydrophobic-like hydration of the protein.[33] Understanding of the nonlinear response of protein solutions has the potential to provide insight into the complexity of hydration water near protein surfaces.

In the current study, lysozyme solutions are investigated with the aim of gaining insight into the nonlinear response of aqueous protein solutions at 0.5 THz. Z-scan measurements reveal a giant nonlinear response of pure water that is ten times greater than in previous literature. This response stems from a persistent thermal lens and is similarly observed for dilute protein solutions. At high protein concentration, the nonlinear response is fully suppressed. These results demonstrate that nonlinear THz driven phenomena are impacted by biological crowding and indicate that the nonlinear properties of protein solutions are correlated to the amount of hydration water.

**Materials and Methods**

*Protein Solutions*

Lysozyme protein from hen egg white (Sigma-Aldrich, purity > 90%) was used as received. Solutions of 2, 4, and 18 mM were prepared by dissolution of an appropriate amount of lysozyme in ultrapure water (resistivity 18.2 MΩ, Merck Milli-Q, Germany). All lysozyme samples were stored at 4°C until further use.

*Terahertz Time Domain Spectroscopy*

Z-scan experiments were carried out at the ELBE center for high power radiation sources with the TELBE facility. The TELBE multicycle undulator was operated with a 50 kHz repetition rate centered at 0.5 THz with about 20% bandwidth FWHM and was utilized as the THz source in a custom-built THz time domain spectroscopy setup as shown in Figure 1. The

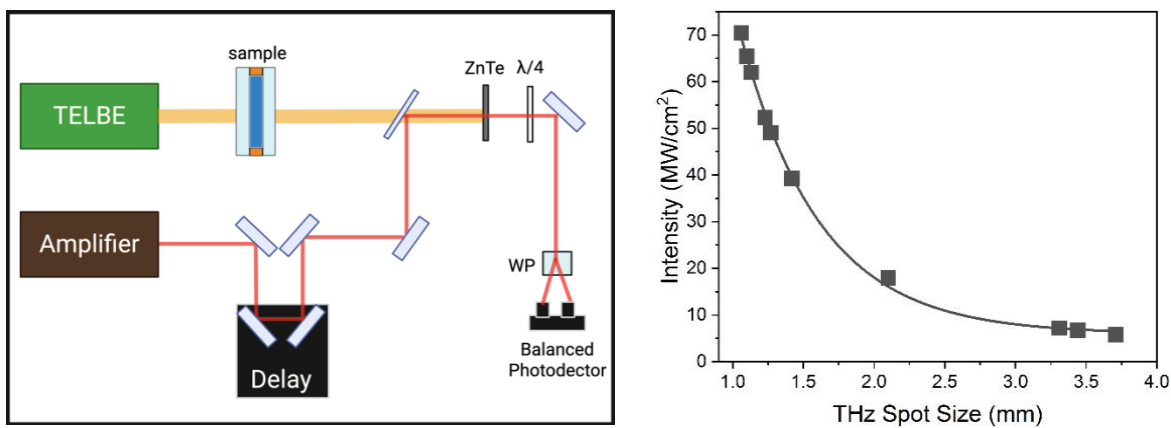

**Figure 1.** (Left) Schematic representation of the Z-scan setup utilizing TELBE as the THz source. (Right) THz beam intensity as a function of the spot size along the z-direction. The line is a guide to the eye.

143 mW THz beam was focused to the sample position (z = 0 mm), where the liquid sample cell with diamond windows (Diamond Materials, Germany) was magnetically mounted to a copper plate that maintained the temperature at 20°C. The THz radiation transmitted through the 50 μm sample was collected, attenuated, and focused to a ZnTe crystal (1 mm thick) for electro-optic sampling. The optical probe pulse came from an 800 nm amplified fs laser (RegA, Coherent, USA) synchronized to the TELBE source. The THz induced change in the polarization of the probe pulse was detected via balanced detection with two silicon photodiodes (Thorlabs, Germany), amplified by a pre-amplifier (SR560, Stanford Research Systems, USA) and processed by a lock-in amplifier (SR830, Stanford Research Systems, USA).

Z-scan measurements were performed by varying the sample position relative to the focus (z = 0 mm) with a manual translation stage, where the furthest position from the focus was z = 18 mm. Figure 1 shows the peak power of the 2.86 μJ THz pulse as a function of the spot size at each z-position. At the focus position, TELBE had a spot size of 1.06 mm, corresponding to a maximum peak power of 70 MW/cm$^2$ (see Supporting Information for details). The peak power decreased away from the focus where linear transmission is expected for low peak powers. Z-scan measurements were conducted on pure water and lysozyme solutions as well as the empty cell for reference. Each measurement was repeated at least three times on two different days to ensure reproducibility.

**Results & Discussion**

The THz electric field transmitted through the empty and water filled cell is shown in Figure 2. The TELBE source is relatively narrowband (~16 ps) and multicycle. Since the electric field is multicycle, the THz electric field trace was Fourier transformed to quantify small changes in the amplitude. The THz field peaked at 0.5 THz (Figure 2, right), and is decreased in the case of the water filled cell due to absorption by the sample. The maximum amplitude of the Fourier transform was used for further analysis. It should be noted that the results obtained below are the same whether the maximum amplitude value or the integrated peak area were used.

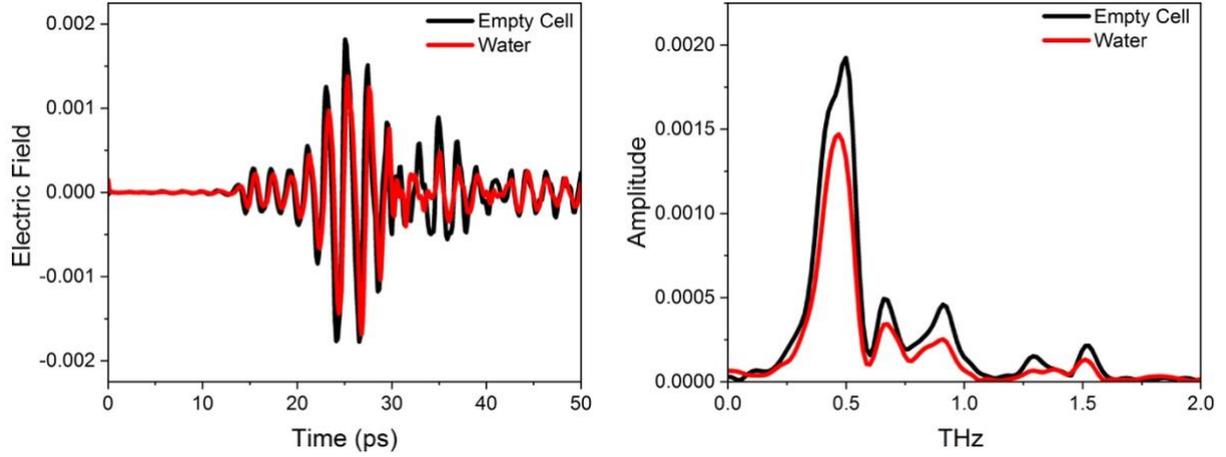

**Figure 2.** (Left) THz electric field transmitted through the empty and water filled cell at z = -2 mm. The THz pulse is multicycle with 8 oscillations and is 16 ps in duration. (Right) Fourier transformed amplitude of the THz pulse transmitted through the empty and water filled cell.

To accurately compare the transmission at different z-positions, the maximum amplitude at each position, A(z), was normalized to the amplitude at the z-position furthest from the focus, A(z = 18 mm). At z-positions far from the focus, the THz intensity is low enough that only linear absorption occurs. Both the sample and the window have a nonlinear THz response near the focus position (see Figure S1 in the Supporting Information (SI)).[34] To isolate the nonlinear transmission of the solution, denoted d$A(z)$, the amplitude ratio of the empty cell ($A_0/A_0$ (z = 18 mm)) is subtracted from that of the water filled cell, as shown in Equation 1.

$$\mathrm{d}A(z) = 1.5 \times \frac{A_\mathrm{sample}(z)}{A_\mathrm{sample}(z = 18~\mathrm{mm})} - \frac{A_0(z)}{A_0(z = 18~\mathrm{mm})} \qquad (1)$$

The factor of 1.5 takes into account that both diamond windows contribute to the response of the empty cell, while 50% of the intensity is absorbed before reaching the second window in the filled cell.[34] The dA(z) for lysozyme protein solutions is shown in Figure 3. Pure water and dilute solutions (2 and 4 mM) show similar behavior, where dA(z) is close to 1 far from the focus, and increases to a value of ~0.7 at the focus. At high concentration, 18 mM, a complete inversion in dA(z) is observed, where values decrease as the z-position approaches the focus, with a minimum dA(z) value of ~0.25. At this concentration, lysozyme begins to have overlapping hydration shells,[35] meaning that bulk water is not readily available. Such results indicate that hydration water has a distinct THz response compared to bulk water, and the origin of this response will be discussed below.

It should be noted that the nonlinear THz transmission of pure water observed here is significantly greater than given in previous literature. A 5% increase in dA was observed in studies with THz fields centered at 0.75 and 1 THz,[29,34] whereas a 30% increase is observed here at 0.5 THz. As water is a highly absorbing material in the THz range, it is possible that heating of the sample, whether homogeneous or inhomogeneous, can cause changes in the real and/or imaginary refractive index, and hence the nonlinear response. The increase in the transmission observed here does not stem from evaporation of water, as the sample remained in the liquid state at all THz intensities. To assess whether THz-induced heating of the sample contributes to the nonlinear transmission measured here, the temperature change in the cell after exposure to the TELBE source was determined using thermal models established in previous literature.[36] The transient temperature in the cell can be estimated from Equation 2,

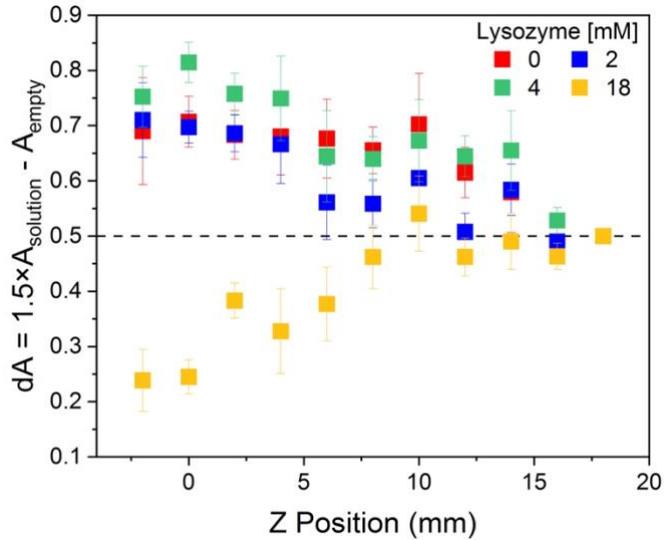

**Figure 3.** The nonlinear transmission of lysozyme solutions at 0.5 THz as a function of z-position. The nonlinear transmission, dA, is determined as the difference in transmission between the filled and empty cell.

$$T(t) = \sum_i \Delta T_i \cdot e^{-t/\tau_{th}} \qquad (2)$$

where $\Delta T_i$ is the steady state temperature increase with the $i$th laser pulse, $t$ is time, and $\tau_{th}$ is the thermalization constant. The thermalization constant is the time scale at which absorbed energy is dissipated from the system as heat, as shown in Equation 3.

$$\tau_{th} = \frac{l(T)^2}{D(T)} \qquad (3)$$

Here $l$ is the penetration depth, which is inversely proportional to the temperature dependent absorption coefficient of water at 0.5 THz, and $D(T)$ is the thermal diffusivity. For water at 20°C, the thermalization constant is ~25 ms.

With the temperature dependent specific heat capacity of water, $C_p(T)$ and the laser pulse energy density $\phi$, $\Delta T_i$ can be determined from Equation 4,

$$\Delta T_i = f(n(T)) \cdot \frac{q \cdot \phi}{l \cdot C_p(T)} \quad (4)$$

where $q$ is the percentage of energy dissipated by contact of the diamond/water interface. Based on previous literature, it is assumed $q = 0.4$. $f(n(T))$ is the temperature dependent Fresnel coefficient, as shown in Equation 5, and describes the intensity of light transmitted by the air/diamond and diamond/water interfaces based on the refractive index of diamond, $n_{dia} = 2.38$, air, $n_{air} = 1$, and water, $n_{water}(T)$. Only the refractive index of water is temperature dependent at 0.5 THz.

$$f(n(T)) = \frac{n_{dia}}{n_{air}} \left( \frac{2n_{air}}{n_{air} + n_{dia}} \right)^2 \frac{n_{water}(T)}{n_{dia}} \left( \frac{2n_{dia}}{n_{air} + n_{water}(T)} \right)^2 \quad (5)$$

To determine the transient temperature of pure water in the sample cell, values of $C_p(T)$, $D(T)$, $n_{water}(T)$, and $\alpha_{water}(T)$ were taken from literature.[37,38] The temperature increase was modeled in steps of 20 μs, equivalent to the TELBE repetition rate. At the focal point (z=0 mm), where the THz is most intense, an increase to ~30°C occurs, as shown in Figure 4 (top panel). Saturation of the transient temperature occurs within ~100 ms following exposure to the THz beam, meaning that the temperature reaches a new equilibrium value quickly. As a time domain scan at a single z-position took ~8 minutes, the temperature increases and stabilizes within the first data point of the scan. The maximum temperature increase relative to 20°C at each z-position is shown in Figure 4 (middle panel). Even at the lowest THz intensity (large z), the temperature in the sample cell increases by ~1°C. To determine whether this temperature increase could explain the observed trend in the nonlinear transmission, a theoretical $dA(z)$ was approximated utilizing the simulated cell temperature, the temperature dependent absorption coefficient, $\alpha(T)$, and the nonlinear transmission of the empty cell (see Figure S1). These results are plotted in Figure 4 (bottom panel) and show an almost linear decrease in $dA(z)$ as the z-position approaches the focus. As the temperature of water increases, the absorption coefficient increases, resulting in decreased transmission. Therefore, uniform thermal heating of the sample does not account for the increased nonlinear transmission observed for pure water.

The theoretical z-scan trace shown in Figure 4 (bottom panel) is similar to the response observed for 18 mM lysozyme (Figure 3), and indicates that thermal heating could contribute in the case of crowded solutions. Equations 2-5 were applied to lysozyme solutions to determine a theoretical $dA(z)$ signal. Values of $C_p(T)$ for lysozyme solutions were taken from literature,[39] while values of $D(T)$ were estimated from extrapolated densities, $\rho(T)$,[40] and thermal conductivities, κ.[41] Values of *n(T)* and *α(T)* were determined from THz-TDS measurements (see Figure S2). Figure 5 shows dA(z) values for an 18 mM lysozyme solution determined from the transient temperature model plotted against the experimental data. The measured values fit fairly well with the model, indicating that uniform sample heating is the origin of the signal measured for 18 mM lysozyme. Solution crowding decreases the heat capacity and leads to lower thermal diffusivity and increased thermalization time constants (~46 ms). As a result, heat accumulates in the measurement cell generating a significantly higher transient temperature (~40°C) (see Figure S3) compared to pure water (~30°C). While dilute lysozyme solutions show a nonlinear response, this effect is annihilated by crowding.

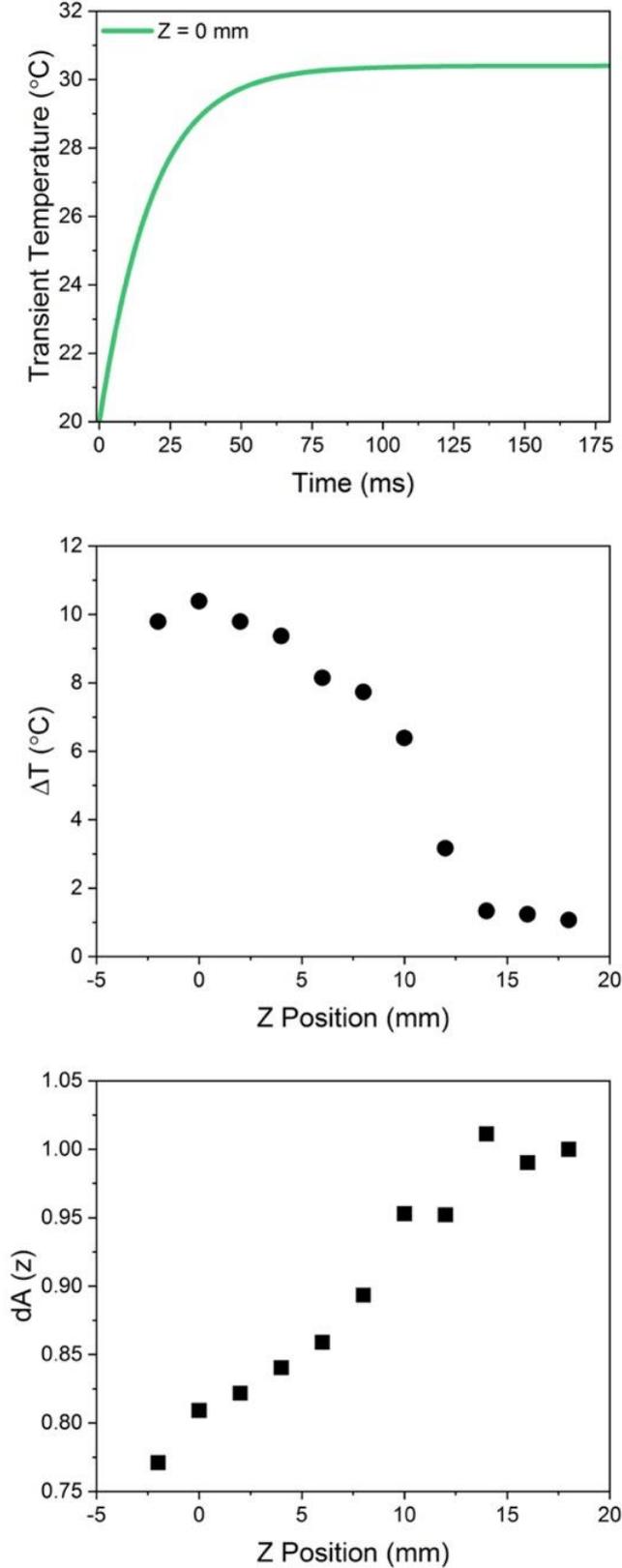

**Figure 4.** (Top) Thermal model of the transient temperature of water in the sample cell at the focus position of the laser beam. (Middle) Transient temperature in the water filled cell relative to 20°C as a function of the z-position. (Bottom) Theoretical transmission determined for the water filled cell,

Based on these results, evaluation of the nonlinear absorption coefficient was limited to dilute lysozyme solutions. In the nonlinear response regime, the total absorption begins to depend on the input laser peak power, *P*, and the linear and nonlinear coefficients both contribute. Considering the Beer Lambert law, the nonlinear absorption coefficient, $\alpha_{NL}$, can be determined from Equation 6,

$$\alpha_{NL} = (0.5 - d_A(z))\frac{\alpha}{P(1-e^{-\alpha L})} \quad (6)$$

where L is the sample thickness. Considering that $L = 50$ μm and that $P = 70$ MW/cm$^2$ at the focus, values of $α_{NL}$ were determined from known $α(T)$ and measured $dA(0)$, and are listed in Table 1. Negative nonlinear absorption values were obtained, indicative of bleaching or saturable absorption. Compared to previous studies, which found that for pure water $α_{NL} = -87$ cm/GW,[29,34] the $α_{NL} = -922 ± 46$ cm/GW obtained here is an order of magnitude larger and $10^7$ greater than values measured in the near infrared and optical regimes.[42,43] Values of $α_{NL}$ for lysozyme solutions vary from that of pure water, indicating that proteins perturb the nonlinear response of the water network.

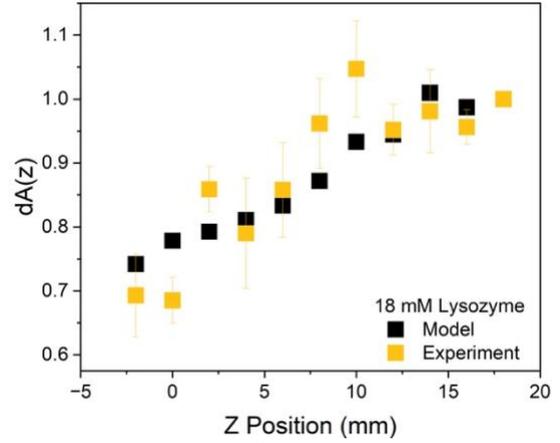

**Figure 5.** Comparison of experimental and theoretical model of transmission of 18 mM lysozyme solution. The model and experiments are in agreement, showing that the 18 mM response results from uniform heating of the sample.

**Table 1.** Nonlinear absorption coefficient for lysozyme solutions determined from the linear absorption coefficients and the nonlinear transmission at the focus position.

| Lysozyme Concentration (mM) | α (cm$^{-1}$) | dA(0) | α$_{NL}$ (cm/GW) |
|---|---|---|---|
| 0 | 193 | 0.707 ± 0.046 | -922 ± 46 |
| 2 | 176 | 0.698 ± 0.029 | -847 ± 29 |
| 4 | 169 | 0.815 ± 0.037 | -1331 ± 31 |

The giant nonlinear THz response reflects a change in material properties, and is indicative of inhomogeneous heat accumulation in the sample.[44] In media where the refractive index is sensitive to temperature changes ($dn/dT$), thermal gradients cause additional focusing or self-defocusing of transmitted light, resulting in altered signal intensity at the detector. The increased nonlinear absorption is consistent with the signature of a self-focusing thermal lens.[45] The dioptric power of a thermal lens is $θ = P_{avg}(1-e^{-α(T)L})(dn/dT)/κπw_0^2$,[46] where $P_{avg}$ is the average power, $dn/dT$ is the thermo-optic coefficient (Figure S2),[37] $κ$ is the thermal conductivity,[38] and $w_0$ is the beam waist. Here, the dioptric power is 357 m$^{-1}$ at z = 0 mm, corresponding to a thermal focal length, $f_{th}$, of 2.80 mm. Typically, thermal lens signals are low for media with high thermal conductivity, such as water. Yet, the $f_{th}$ determined here has the same order of magnitude as the Rayleigh range, $z_r = πw_0^2/λ = 1.47$ mm, indicative of a strong thermal lensing effect. Assuming moderate thermal lensing occurs at a dioptric power of ~10, where $f_{th} > z_r$ by a few orders of magnitude, a maximum beam waist of of ~3.2 mm is required to induced the thermal lensing effect. This is reflected in the nonlinear transmission datat shown in Figure 3, and indicates that thermal lensing contributes to the measured effect.

The rise time of a thermal lens depends on the acoustic transit time, defined as $r = d/s$, where $d$ is the THz spot size and $s$ is the speed of sound in the media.[45,47] Considering the spot size was $d = 1.06$ mm and $s = 1500$ m/s for water, it is estimated that the rise time of a thermal lens in the liquid cell is ~0.7 µs. Relaxation of a thermal lens is governed by the thermalization constant, which is ~19 ms for water at 30°C. As the TELBE repetition rate is 50 kHz, i.e. 20 µs between pulses, there is practically no relaxation of the thermal lens between successive pulses. Previous studies of the nonlinear absorption of water used much slower repetition rates of 50 Hz,[34] for which there would be sufficient time for thermal relaxation between pulses, or used a liquid jet as the sample environment,[29] where thermal gradients would be partially washed away between successive pulses.[36] Therefore, it is unlikely that thermal lensing substantially contributed to signals measured in previous studies. The nonlinear response determined here comes from the generation of a persistent thermal lens that causes additional focusing, such that the effective length shifts to ~2.70 mm. The smaller nonlinear absorption values obtained for lysozyme solutions is also consistent with thermal lensing. Thermal conductivities of dilute lysozyme solutions are greater than or equal to that of pure water,[41] while the absorption coefficient at 0.5 THz decreases with increasing lysozyme concentration (see Table 1). Therefore, the strength the thermal lensing effect is expected to be weaker for lysozyme solutions and is reflected in smaller nonlinear absorption coefficients obtained. Such results demonstrate that THz nonlinearities are sensitive to solutes in aqueous systems.

**Conclusions**

Here, a giant nonlinear transmission is observed for pure water and dilute lysozyme solutions. This nonlinear response is completely suppressed for crowded lysozyme solutions, and stems from uniform heating of the sample, where decreased heat capacity contributes. In the case of dilute solutions, the response comes from the formation of a persistent thermal lens, driven by the intense electric field and high repetition rate of the TELBE source. As smaller nonlinear absorption coefficients were determined for lysozyme solutions, and no nonlinear response was observed for the crowded solution, these results suggest that the nonlinear response is related to the anisotropy of bulk water available in solution. To uncover how the nonlinear response of hydration water compares to bulk water, the use of THz sources with lower repetition rates or liquid jets would be required to overcome thermal heating effects.

**Author Contributions**

E.M.A., I.I., and J.D. designed research. All authors performed research. E.M.A. analyzed data. All authors participated in the writing of the manuscript.

**Conflict of Interest**

The authors declare no conflict of interest.

**Acknowledgements**

E.M.A. acknowledges financial support by Germany's Excellence Strategy-EXC 2068—Projektnummer 390729961 from the Deutsche Forschungsgemeinschaft (DFG, German Research Foundation) via the Cluster of Excellence Physics of Life. Parts of this research were carried out at ELBE at the Helmholtz-Zentrum Dresden-Rossendorf e.V., a member of the Helmholtz Association. We thank A. Czajkowski and Q. Minh Thai for experimental support and F. Novelli for discussion.